\documentclass[twocolumn,prl,a4,aps,showpacs,footinbib,amsmath,amssymb,superscriptaddress]{revtex4-1}
\usepackage{graphicx}
\usepackage{dcolumn}
\usepackage{color}
\usepackage{float}
\usepackage{bm}
\usepackage{natbib,hyperref}
\usepackage{amsmath,amssymb,amsfonts}
\usepackage[normalem]{ulem}

\newcommand{\beq}{\begin{equation}}
\newcommand{\eeq}{\end{equation}}
\newcommand{\beqa}{\begin{eqnarray}}
\newcommand{\eeqa}{\end{eqnarray}}

\begin{document}

\title{Plumbene on a magnetic substrate: a combined STM and DFT study}

\author{Gustav Bihlmayer} \email{g.bihlmayer@fz-juelich.de}
	\affiliation{Peter Gr\"{u}nberg Institut and  Institute for Advanced Simulation,
                        Forschungszentrum J\"{u}lich \& JARA, 52425 J\"{u}lich, Germany}
\author{Jonas  Sassmannshausen}
       \affiliation{Department of Physics, University of Hamburg, D-20355 Hamburg, Germany}
\author{Andr\'{e}  Kubetzka}
       \affiliation{Department of Physics, University of Hamburg, D-20355 Hamburg, Germany}
\author{Stefan Bl\"{u}gel}
	\affiliation{Peter Gr\"{u}nberg Institut and  Institute for Advanced Simulation,
                        Forschungszentrum J\"{u}lich \& JARA, 52425 J\"{u}lich, Germany}
\author{Kirsten  von Bergmann} \email{kbergman@physnet.uni-hamburg.de}
       \affiliation{Department of Physics, University of Hamburg, D-20355 Hamburg, Germany}
\author{Roland  Wiesendanger}
       \affiliation{Department of Physics, University of Hamburg, D-20355 Hamburg, Germany}

\date{\today}


\begin{abstract}
As heavy analog of graphene, plumbene is a two-dimensional material with strong spin-orbit coupling effects. Using scanning  tunneling microscopy (STM), we observe that Pb
forms a flat honeycomb lattice on an Fe monolayer on Ir(111). In contrast, without the Fe layer, a $c(2\times4)$ structure of Pb on Ir(111) is found. We use density
functional theory (DFT) calculations to rationalize these findings and analyze the impact of the hybridization on the plumbene band structure. In the unoccupied states
the splitting of the Dirac cone by spin-orbit interaction is clearly observed while in the occupied states of the freestanding plumbene we find a band inversion that
leads to the formation of a topologically non-trivial gap. Exchange splitting as mediated by the strong hybridization with the Fe layer drives  a quantum spin Hall to
quantum anomalous Hall state transition.
\end{abstract}

\maketitle


Since the exotic properties of graphene have been discovered about 15 years ago~\cite{Novoselov:04.1}, the field of two-dimensional materials in general and honeycomb structures in particular
has seen a dramatic increase in popularity~\cite{Neto:09.1}. Topological properties in these lattices depend significantly on the strength of spin-orbit coupling (SOC)
effects that are notoriously small in graphene, especially at the $\overline{\rm K}$ point~\cite{Liu:10.1}. Therefore, although the quantum spin Hall effect (QSHE) was first
theoretically predicted for graphene~\cite{Kane:05.1}, experimentally it was first verified in materials containing heavy elements like HgTe quantum
wells~\cite{Koenig:07.1}.  Since then, numerous studies have focused on the synthesis and properties of heavier analogs of graphene like silicene~\cite{Molle:18.1},
germanene~\cite{Zhang:16.1} or stanene (Sn)~\cite{Zhu:15.1}. But the formation of double bonds in this series seems to be restricted to the carbon-based material only and
freestanding heavier analogs are considered unlikely to form~\cite{Hoffmann:13.1}. Consequently, the first silicene was reported as adlayer on Ag(111)~\cite{Vogt:12.1}
and it is still challenging to balance the interaction with the substrate required for formation with the electronic independence necessary to study the topological
properties via electronic transport effects~\cite{Molle:18.1}. Furthermore, all heavier analogs of graphene have a tendency to pronounced buckling of their honeycomb
structures, resulting in severe changes of the electronic properties as compared to the ideal flat structures~\cite{Huang:14.1}. Therefore, it came recently as a welcome
surprise that  stanene was observed to grow on Cu(111) as a flat honeycomb lattice~\cite{Deng:18.1}. Despite the metallic substrate, a topological edge state could be
observed on these islands - although 1.3~eV below the Fermi level.

In this quest for heavy honeycomb structures the Pb analog, plumbene, appeared relatively late on the scientific stage. Isoelectronic in its valence shell with C, Si, Ge
and Sn it is the heaviest graphene analog and expected to show the most pronounced SOC effects~\cite{Huang:14.1}. Density functional theory (DFT) studies of plumbene
predicted the formation of a buckled honeycomb structure but without band inversion near the Fermi level~\cite{Yu:17.1}. Electronically it is similar to a Bi(111) bilayer 
with less electrons ($Z_{\rm Bi}=83$, $Z_{\rm Pb}=82$). DFT studies suggested that doping or chemical modification of plumbene~\cite{Zhao:16.1} might be necessary to
achieve topological effects. Maybe it is because of these findings that the quest for plumbene has not really started yet.

Using scanning tunneling microscopy (STM) and DFT we show in this Letter that (i) using an appropriate substrate it is possible to form a flat plumbene lattice  
and (ii) that the electronic properties of 'flat plumbene' are rather exciting: Calculations predict  a band-inversion in the valence bands that leads to 
topologically protected edge states. Further (iii), on the ferromagnetic substrate that enables the formation of plumbene the induced exchange splitting 
drives this feature into a quantum anomalous Hall gap. Such exchange coupling opens the way to realize a quantum Hall effect without external magnetic 
field, a phenomenon envisioned theoretically in the eighties~\cite{Haldane:88.1} and only recently realized experimentally~\cite{Chang:13.1} at very low temperatures.

\begin{figure}[t]
\centering
\includegraphics[width=1\columnwidth]{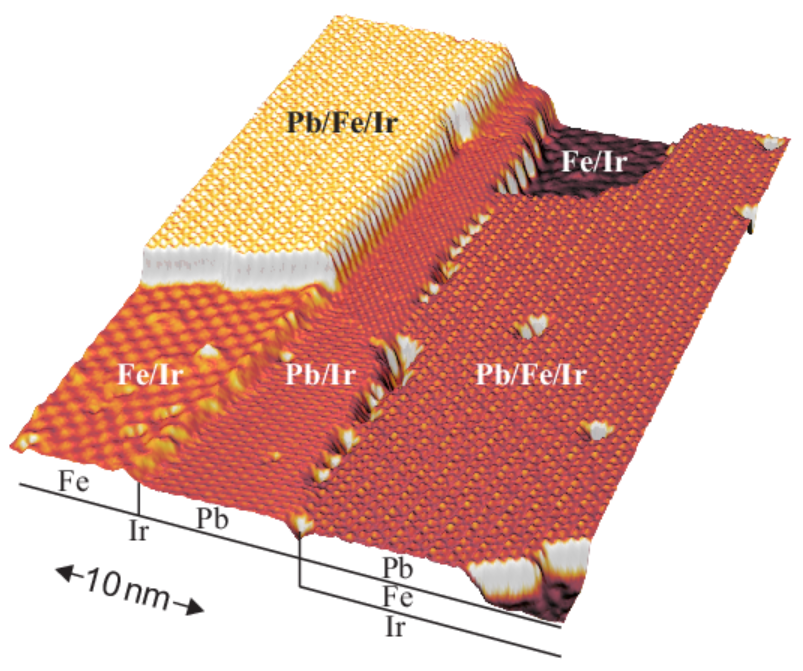}
\caption{Pseudo 3-dimensional STM image of a sample of Pb on Fe on Ir(111), the different exposed surfaces are labeled. Pb was deposited well below room temperature and 
grows as monolayer high ordered islands both on the bare Ir surface and on the Fe monolayer on Ir. The superstructure on the pseudomorphic Fe monolayer on Ir is not 
due to the structure but corresponds to the magnetic signal of the nanoskyrmion lattice. Measurement parameters: $U=+5$\,mV, $I=3$\,nA, $T=4$\,K, Cr-bulk tip.
\label{fig:overview}
}
\end{figure}

We have deposited sub-monolayer amounts of Pb onto a sample  with extended Fe monolayer areas on an Ir(111) single crystal surface, see overview STM image in
Fig.~\ref{fig:overview}. Because we observed severe intermixing of Pb and Fe for Pb growth at room temperature we have cooled the Fe/Ir(111) to about $140$~K prior to the Pb 
deposition~\cite{SassmannPRB2018}, which results in large and well-ordered patches of Pb both on the bare Ir(111) and the Fe-covered Ir(111), see labels for the different 
layers in Fig.~\ref{fig:overview}. The Fe monolayer grows pseudomorphically in fcc stacking on the Ir(111) substrate and in this spin-polarized STM measurement~\cite{Wiesendanger2009} 
the observed roughly square superstructure with a periodicity of about 1\,nm originates from the magnetic nanoskyrmion lattice~\cite{Heinze2011}. No magnetic signal has 
been observed on the Pb monolayers.

\begin{figure}[t]
\centering
\includegraphics[width=1\columnwidth]{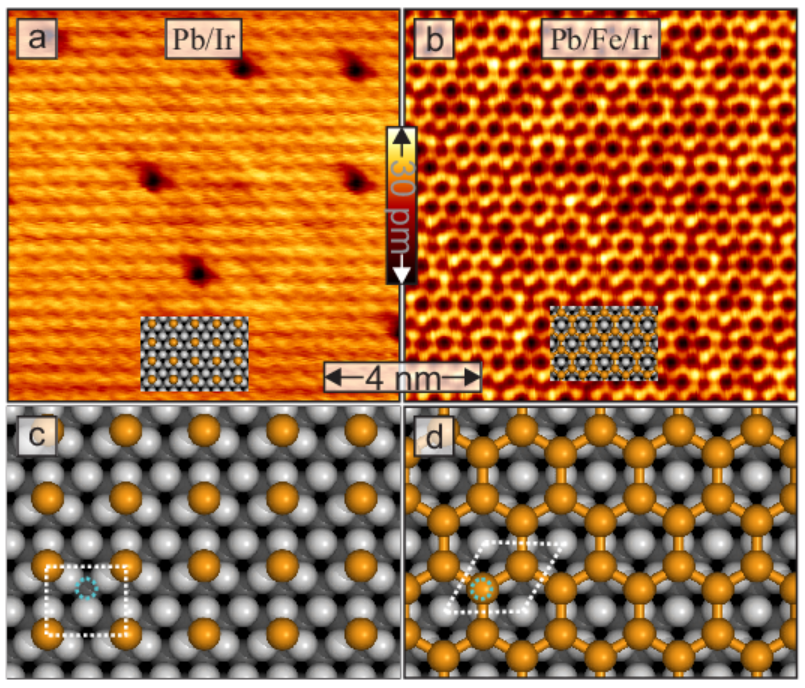}
\caption{Closer view constant-current STM images of the Pb monolayer on (a) the bare Ir(111) surface and (b) the Fe monolayer on Ir(111). Both the symmetry 
and the atomic distances for the two differently ordered Pb monolayers are different and the corresponding structure models are presented in (c) and (d). 
The white dashed rectangle in (c) refers to the primitive unit cell of the $c(2\times4)$ structure and the white dashed diamond in (d) marks the $p(2\times2)$ unit cell; blue 
dashed circles mark positions that are empty for a lower atom-density configuration and occupied for a higher-atom density configuration (confer text). 
Measurement parameters: (a) $U=+10$\,mV, $I=1.5$\,nA; (b) $U=+5$\,mV, $I=2$\,nA; both $T=4$\,K, Cr-bulk tip.
Information on the bias dependence is provided in the supplementary information, Fig.~S1.
\label{fig:structure} }
\end{figure}

A closer view on the Pb  deposited directly on the Ir and on Fe/Ir is shown in Figs.~\ref{fig:structure}(a) and (b), respectively. Here the superstructures are of structural origin. Due to the large lattice mismatch of Pb and Ir, the Pb overlayers are not pseudomorphic, but instead form layers with reduced atom density. Nevertheless, the Pb superstructures are commensurate with the substrate and atoms reside at specific adsorption sites of the Ir or Fe/Ir, an indication that the Pb-substrate interaction is comparable to the Pb-Pb interaction.

On Ir(111) we find a $c(4\times2)$ Pb overlayer that can be described with a rectangular 
unit cell, see white dashed 
rectangle in Fig.~\ref{fig:structure}(c), like in the graphene intercalated Pb films on Ir~\cite{Calleja:15.1}. The Pb atoms all occupy the same adsorption sites and the Pb-Pb distances are 4.70\,\AA\ and 5.43\,\AA\ for the two orthogonal directions. For symmetry reasons three rotational domains are found on a larger scale.

In contrast, on the Pb/Fe/Ir we find a honeycomb arrangement of the Pb atoms, see experimental data and corresponding structure model in Figs.~\ref{fig:structure}(b),(d), i.e.\ the Pb grows on Fe/Ir(111) in the modification of plumbene. The structural $p(2\times2)$ unit cell (see white dashed diamond) contains two Pb atoms that adsorb in a fcc and a hcp site. The Pb atom density of this honeycomb plumbene layer is twice that of the $c(4\times2)$ Pb overlayer on Ir.
Since both the Fe/Ir and the Ir(111) surface have the same symmetry and identical atomic distances, the difference in the Pb overlayer structures cannot originate from geometrical reasons.
While the structural unit cell of the honeycomb has lattice vectors with a length of 5.43\,\AA, the Pb-Pb distance is only
$\sqrt{6}/3 \mathrm{a}_\mathrm{Ir} = 3.13$~\AA, i.e.\  more than 10\% shorter than in fcc Pb ($\mathrm{a}_\mathrm{Pb}/\sqrt{2} = 3.52$~\AA). Comparing graphene with
diamond a similar contraction of bond distances (1.42~\AA~ vs.\ 1.55~\AA) can be observed, suggesting that the Pb-Pb bond is modified in a similar way as the C-C bond for the two different 
allotropes.

Using density functional theory we study the energetics of different Pb monolayers on Ir(111) and Fe/Ir(111). Based on the experimental findings we compare four possible arrangements of Pb at these surfaces: $p(2\times2)$
and $c(2\times4)$ unit cells (uc.) with one or two Pb atoms per cell, i.e.\ the structures shown in Figs.\,\ref{fig:structure}(c) and (d) with and without the atoms at 
positions marked by the dashed blue circle. 

Comparing fcc and hcp adsorption sites, a single Pb atom on Ir(111) forming a $p(2\times2)$ or $c(2\times4)$ structure always prefers the fcc site by 49 meV/Pb. 
The $p(2\times2)$ and $c(2\times4)$ arrangements differ by only 5 meV in favor of the former. The fact that experimentally a $c(2\times4)$ structure is observed might be
related to neglected effects from the vibrational entropy or limitations of the computational method. To put two Pb atoms into a $c(2\times4)$ uc., however, requires 54
meV/Pb more energy than a single one and the formation of a Pb honeycomb lattice in the $p(2\times2)$ cell is energetically 216 meV/Pb more expensive than the
$c(2\times4)$ arrangement with the same atom density. Thus, on Ir(111) a low Pb atom density with larger Pb-Pb distances is energetically favorable as found in the
experiments reported here and in the graphene covered system~\cite{Calleja:15.1}. Our STM simulations based on the local density of states (LDOS) also show good
agreement with the experimental images (see supplementary information, Fig.~S2).

\begin{figure}[t]
\centering
\includegraphics[width=1\columnwidth]{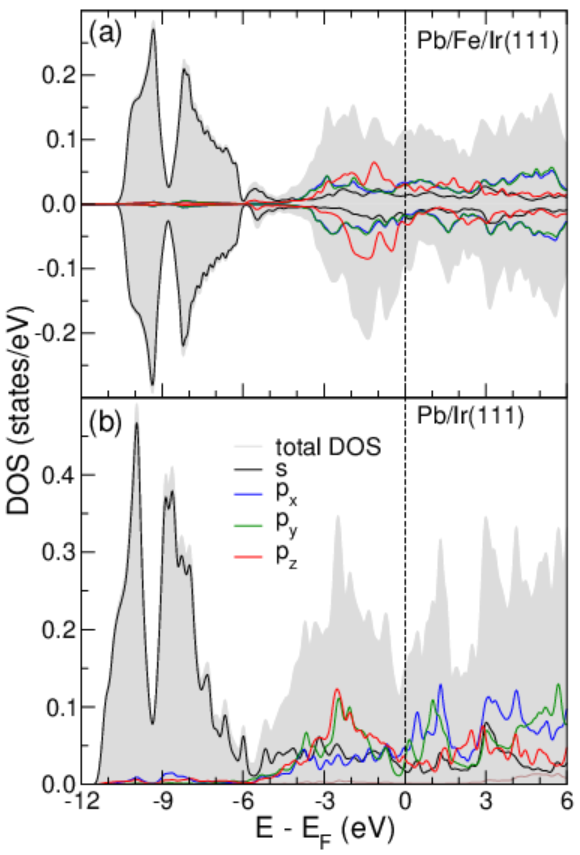}
\caption{(a) Locally and orbitally resolved DOS of the Pb atoms in the honeycomb structure on Fe/Ir(111). The total Pb DOS is shown as gray background, the $s$, $p_x$,
$p_y$ and $p_z$ contributions are shown in black, blue, green and red, respectively. Positive and negative values correspond to majority and minority spins, respectively. (b) The same quantities for the Pb honeycomb lattice on Ir(111). \label{fig:LDOS} }
\end{figure}

This high energy cost to form a Pb honeycomb structure on Ir(111) is contrasted by the energetics of Pb on Fe/Ir(111): here the honeycomb lattice is favored 
by 11 meV/Pb atom over the $c(2\times4)$ structure with the same areal Pb density (2 Pb/uc.) and by 44 meV over a $p(2\times2)$ arrangement with only one Pb atom per unit cell. 
The structural relaxation shows that the honeycomb layer is almost completely flat with a corrugation of 0.002~\AA.
In all cases we assumed a ferromagnetic order of the Fe layer. Test calculations of Fe layers with antiferromagnetic nearest neighbor interactions 
lead to much higher total energies (see supplementary information). 
A similar magnetic hardening was observed when coronene is adsorbed on the Fe/Ir(111) nanoskyrmion lattice~\cite{Brede:14.1} and can occur generally, when $p$ electrons of molecules interact with a magnetic layer underneath~\cite{Friedrich:15.1}.

To understand the radically different energetics of Pb on Ir(111) and on Fe/Ir(111) we analyze the orbitally resolved density of states (DOS) for the honeycomb structure
in Fig.~\ref{fig:LDOS}. In the case of the Fe/Ir(111) substrate, we see that the DOS of the Pb $p_x$ and $p_y$ states is almost degenerate and rather featureless 
near the Fermi level, while the $p_z$ states show characteristic peaks at or below 1~eV binding energy where the minority Fe states are located (the majority states 
of Fe are peaked between 2 and 3~eV binding energy, see supplementary information, Fig.~S3). While in this case the in-plane and out-of-plane oriented $p$ orbitals of Pb seem rather decoupled 
(as expected from a 2D structure bonded by  $p_z$ orbitals to the substrate), in the Pb/Ir(111) case we find a completely different orbital arrangement with $p_y$/$p_z$ 
orbitals bonding to the Ir substrate. Although also in this case an almost flat Pb  honeycomb structure is obtained, the involved orbitals are completely different. 
Therefore, it can be concluded that the hybridization with the Fe states, that are available in the energy range of the Pb $p_z$ states, is responsible 
for the formation of the 2D Pb honeycomb lattice.

\begin{figure}[t]
\centering
\includegraphics[width=1\columnwidth]{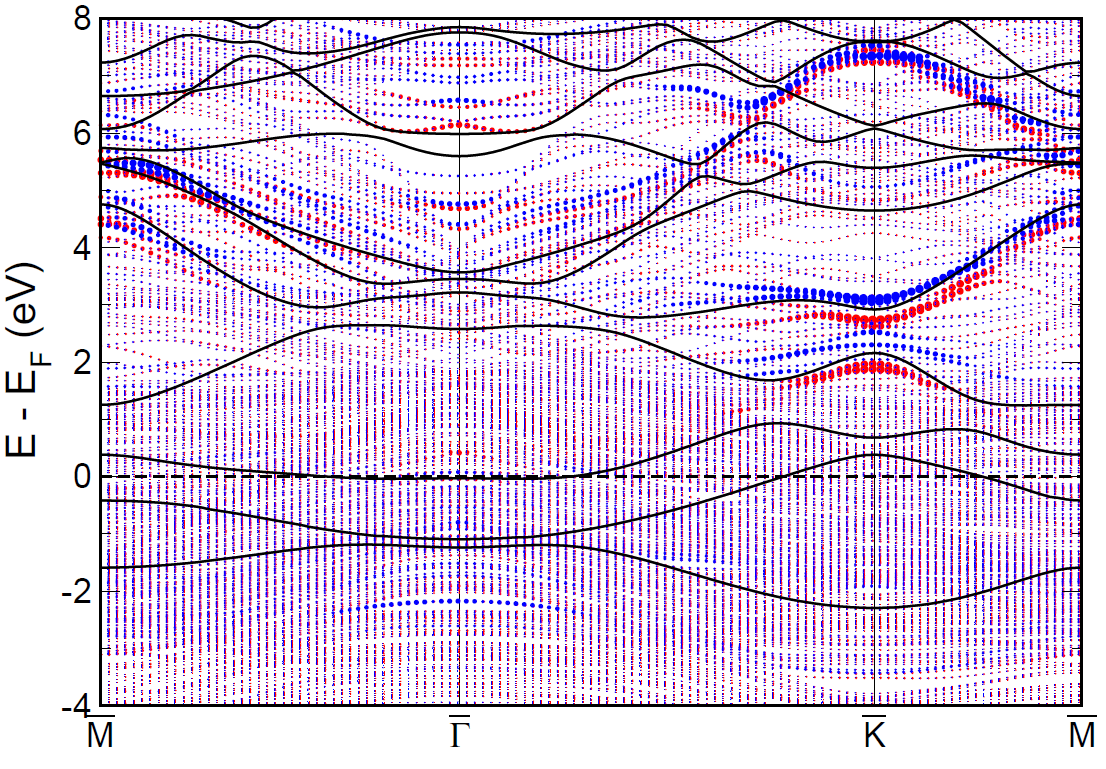}
\caption{Red/blue: spin polarization of the Pb states of a Pb honeycomb structure on Fe/Ir(111). Spin-orbit coupling is included and the spin-quantization axis is assumed to be normal to the surface. In black the band structure of a freestanding plumbene layer with the same structure as the deposited one is shown. \label{fig:band_SOC} }
\end{figure}

Turning now to the band structure of Pb on Fe/Ir(111), it is instructive
to investigate first the plumbene layer without the substrate (black lines in Fig.~\ref{fig:band_SOC}). Below the Fermi level two rather flat bands 
can be found and the $p_z$ states form a hole pocket around the $\overline{\rm K}$ point,
compensated by a shallow electron pocket of the antibonding $p_{x,y}$ states at $\overline{\Gamma}$. At around 1 eV binding energy the bonding $p_{x,y}$ states show a
band inversion with the $p_z$ states at the $\overline{\Gamma}$ point, very similar to the band inversion observed for stanene in Ref.~\cite{Deng:18.1}. On the metallic
substrate these Pb states are, however, strongly hybridized with the Fe $d$ states and --as evident from Fig.~\ref{fig:band_SOC}-- only above the Fermi level the
individual Pb states can be identified again. The induced spin-splitting in these states is substantial, between 0.2 and 0.4 eV. This can be seen best at the $\overline{\rm K}$ 
point, where the Dirac-type band crossing is lifted by SOC (for the band structure without SOC see supplementary information, Fig.~S4.

\begin{figure}[t]
\centering
\includegraphics[width=1\columnwidth]{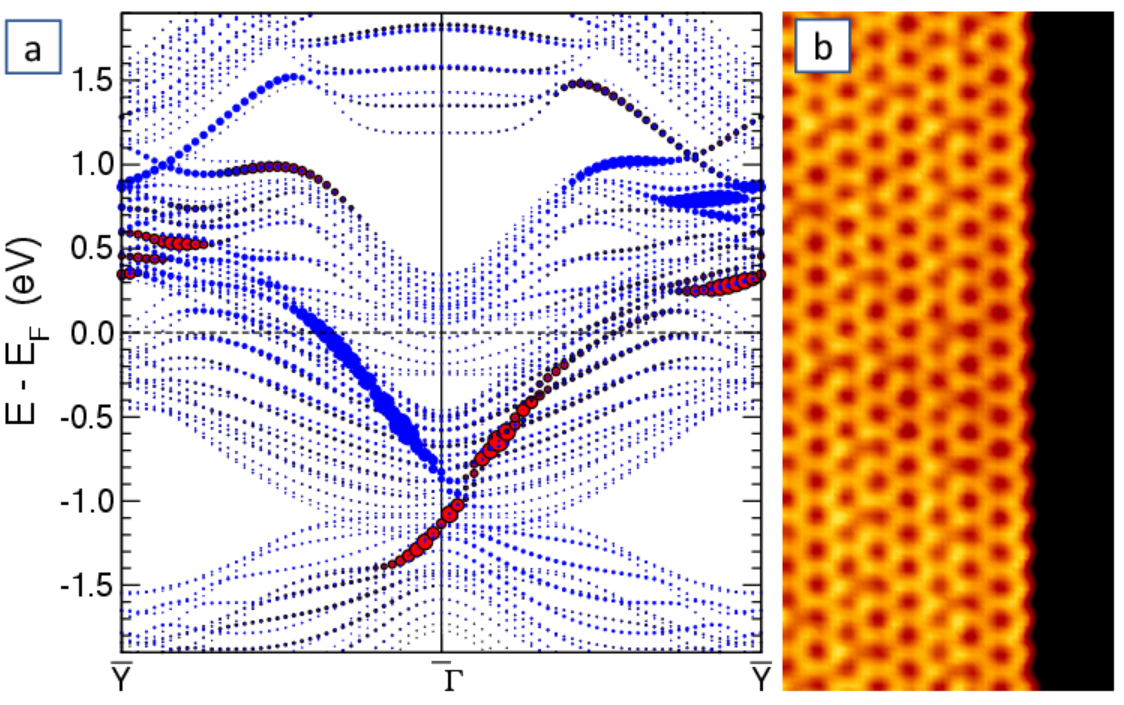}
\caption{(a) Band structure of an edge-hydrogenated plumbene ribbon in an external magnetic field. The size of the dots marks the weight of the states on one of the zig-zag
edges, the color indicates the spin character. (b) Constant-current STM image of the typical zigzag configuration of the edge of a plumbene island on Fe/Ir; measurement 
parameters: $U=-10$\,mV, $I=2$\,nA, $T=4$\,K, Cr-bulk tip. \label{fig:QAHS} }
\end{figure}

Although the features of the Pb bands are blurred by hybridization, it is of interest to study the consequences of the above-mentioned band inversion of plumbene at the
$\overline{\Gamma}$ point. Strong SOC in Pb opens a 150 meV band gap between the states that hosts topologically protected edge states. In the calculation of the
edge-hydrogenated plumbene zig-zag ribbon the linear dispersion of the edge state is clearly visible in the gap (see supplementary information, Fig.~S5). This is a clear signature of the quantum spin
Hall effect (although in the occupied states) as expected from the parity analysis of the bands. Since the Fe substrate induces a strong exchange field in the plumbene,
we can further look for the appearance of a quantum anomalous Hall gap in the system. To simulate the effect of the exchange field in the unsupported  plumbene ribbon, we
add an external magnetic field in the calculation that reproduces the spin-splitting of the Pb bands seen in Fig.~\ref{fig:band_SOC}. Although in the projection on the
edge the spin-split band-edges now overlap, the single, spin-polarized edge channel is clearly visible in this system (Fig.~\ref{fig:QAHS}).

We have shown experimental and theoretical evidence that the heaviest member of the graphene family, plumbene, can be prepared on an Fe layer on Ir(111). 
Selective hybridization of the Pb $p_z$ states with the Fe minority $d$ states stabilize the honeycomb structure, while on bare Ir(111) a rectangular 
$c(2 \times 4)$ Pb layer is formed. The ferromagnetic substrate induces a significant exchange splitting in the plumbene. Freestanding plumbene with this
lattice constant shows a band-inversion in the occupied states and topologically protected edge states. With the  exchange splitting this quantum-spin-Hall gap
is changed into a quantum-anomalous-Hall gap providing a protected charge channel. Although on the metallic substrate these states will be considerably 
broadened and not accessible to transport measurements, with a modified substrate this system might be a nice platform to study the properties of edge states
of a Chern insulator.

\section*{Acknowledgments}

We would like to thank Niklas Romming for technical assistance with the experiments. We gratefully acknowledge the Collaborative
Research Center SPP 1666 of the Deutsche Forschungsgemeinschaft (DFG) and the computing time on the JURECA  supercomputer of the Jülich Supercomputing Centre (JSC). Financial support from the ERC Advanced Grant ADMIRE is gratefully acknowledged.

\bibliography{PbIr}

\end{document}